\documentclass[useAMS,usenatbib]{mn2e}

\usepackage{amsmath}
\usepackage{epsfig}
\usepackage{graphicx}
\newcommand{\parenthnewln}{\right.\\&\left.\quad\quad{}}

\title[Relativistic effects in the chaotic Sitnikov problem]{Relativistic effects in the chaotic Sitnikov problem}
\author[T. Kov\'acs, Gy. Bene, and T. T\'el]{T. Kov\'acs$^{1,2},$\thanks{E-mail:
t.kovacs@astro.elte.hu;} Gy. Bene$^{3},$ and T. T\'el$^{3}$\\
$^{1}$Max Planck Institute for the Physics of Complex Systems, D-01187 Dresden,Germany\\
$^{2}$Konkoly Observatory of the Hungarian Academy of Sciences, P.O. Box 67, H-1525 Budapest, Hungary\\
$^{3}$Department of Theoretical Physics, E\"otv\"os University, H-1117 Budapest,Hungary}
\begin{document}

\date{Accepted 1988 December 15. Received 1988 December 14; in original form 1988 October 11}

\pagerange{\pageref{firstpage}--\pageref{lastpage}} \pubyear{2009}

\maketitle

\label{firstpage}

\begin{abstract}
We investigate the phase space structure of the relativistic Sitnikov problem in the first post-Newtonian approximation. The phase space portraits show a strong dependence on the gravitational radius which describes the strength of the relativistic pericentre advance. Bifurcations appearing at increasing the gravitational radius are presented. Transient chaotic behavior related to escapes from the primaries are also studied. Finally, the numerically determined chaotic saddle is investigated in the context of hyperbolic and non-hyperbolic dynamics as a function of the gravitational radius.  
\end{abstract}

\begin{keywords}
methods: numerical, chaos, relativistic processes, scattering, celestial mechanics
\end{keywords}

\section{Introduction}

The Sitnikov problem (SP) \citep{Sit} is one of the simplest dynamical systems in celestial mechanics that provides all kinds of chaotic behavior. The configuration of the system is defined by: two point-like bodies of equal masses (called primaries) orbiting around their common centre of mass due to their mutual gravitational forces, and a third body of negligible mass moving along a line, perpendicular to the orbital plane of the primaries, going through their barycentre. For the circular motion of the primaries, the problem is integrable and \citet{Mac} gave a closed form analytical solution with elliptic integrals. \citet{Mos} showed the existence of chaotic behavior using symbolic dynamics. 

In the last decades the problem was investigated in details both analytically and numerically. \citet{Liu} derived a mapping model to investigate the problem. \citet{Wod} introduced a new formulation for the equation of motion by using the true anomaly of the primaries as independent variable. \citet{Hag} extended the analytical approximations up to very high orders by using extensive computer algebra. \citet{Dvo1} showed by numerical computations that invariant curves exist for small oscillations around the barycentre. \citet{Alf} determined invariant rotational curves by applying the Birkhoff normal form of an area-preserving mapping. Periodic solutions were studied by \citet{Per}, \citet{Bel}, \citet{Jal}, \citet{Kal}, and \citet{Cor}. The complete phase space was studied numerically by \citet{Dvo2} and \citet{Kov1}.

The Sitnikov problem is the perfect manifestation of a scattering process in which a particle approaches a dynamical system from infinity, interacts with the system and ultimately the particle leaves it. Escapes to infinity in Sitnikov problem were studied by \citet{Kov2}. The test particle can escape the system via different exits, thus one can determine the basins of escape, since in these systems infinity acts as an attractor for an escaping particle. A detailed study about basins of escape can be found in \citet{Ble1} and \citet{Con1}.

An interesting question is how the structure of the phase space changes due to relativistic effects. Since \citet{Rob} gave the solution of the relativistic two-body problem in the post-Newtonian (PN) approximation, many papers have dealt with pericentre advance in celestial mechanics, especially in the case of binary pulsars where the masses of celestial bodies are of the same magnitude \citep{Dam1,Wag}. It is an established fact that the pericentre precession in the post-Newtonian two-body problem is the same as in Schwarzschild's metric \citep{Lan,Dam2} expressed by the total mass. Namely, the pericentre advance in one revolution is
\begin{equation}
\Delta\phi = 6\pi\frac{k(m_{1}+m_{2})}{ac^{2}(1-e^{2})},
\label{eq:dphi}
\end{equation}
where $k$ represents the gravitational constant, $m_1$ and $m_2$ are the masses of the bodies, and $a$ and $e$ describe the classical semi-major axis and the eccentricity, respectively. In Eq.~(\ref{eq:dphi}) $c$ denotes the speed of light. 

The aim of the present work is to show the relativistic dynamics of the Sitnikov problem by taking into account the leading post-Newtonian ''perturbation''. The paper is organized as follows. In Sec. 2 we describe the model: first, the post-Newtonian two-body problem, then the relativistic Sitnikov problem. Section 3 contains our numerical results. Sec. 3.1 concentrates on the phase space structure of the relativistic SP, and 3.2 deals with chaotic scattering. In Section 4 we draw our conclusions. The derivation of the equations of motion can be found in the Appendix.

\section[]{Description of the model}
The Sitnikov problem is a particular case of the restricted three-body problem. The third, mass-less body has no effect on the primaries' motion, neither in the classical, nor in the relativistic case. Therefore, the problem can be split into two parts. The solution of the two-body problem is needed for the determination of the motion of the test-particle. The instantaneous position of primaries provides the time dependent driving acting on the test-particle. 

\subsection{Post-Newtonian two-body problem}
All our results are given in the leading post-Newtonian (PN) approximation based on an assumption of weak inter-body gravitational field and slow orbital motions. Beyond the classical limit, it contains terms of order $v^{2}/c^{2},$ where $v$ is a typical orbital velocity and $c$ is the speed of light \citep{Cal}.

Throughout the paper the length unit will be chosen as the semi-major axis of the classical two-body problem: $a=-km_{1}m_{2}/(2E_{c}),$ where $E_{c}$ is the classical energy in the centre of mass frame. The time unit is taken \textbf{as} $T=a^{3/2}[k(m_{1}+m_{2})]^{-1/2}$ from Kepler's third law. The energy unit will be $km_{1}m_{2}/a.$ The total dimensionless classical energy becomes $E_{c}=-0.5.$

The equations of the relative motion in the post-Newtonian centre of mass frame can be derived from the Lagrangian given in \citet{Dam2}. The equation for the relative coordinate $\mathbf{r}$ reads in dimensionless form up to first order in $\lambda$
\begin{equation}
  \begin{split}
    \mathbf{\ddot r}=\mathbf{\dot v}=-\frac{\mathbf{r}}{r^{3}}&
    +\lambda\left[-(1+3\nu)\frac{\mathbf{r}}{r^{3}}v^{2}+\frac{3}{2}\nu\frac{\mathbf{r}}{r^{5}}(\mathbf{rv})^{2}\parenthnewln
      +(4+2\nu)\frac{\mathbf{r}}{r^{4}}+(4-2\nu)\frac{\mathbf{v}}{r^{3}}(\mathbf{rv})\right],
  \end{split}
  \label{eq:pn}
\end{equation}
where $\nu=m_{1}m_{2}/(m_{1}+m_{2})^{2}$ is the effective mass and 
\begin{equation}
  \lambda=k(m_{1}+m_{2})/ac^{2}
  \label{eq:lambda}
\end{equation}
denotes the dimensionless \textit{gravitational radius} with $a$ as the classical semi-major axis.

The invariance of the Lagrangian under time translation and spatial rotation implies the existence of four first integrals, the energy and the angular momentum of the binary system in the centre of mass frame:
\begin{equation}
  \begin{split}
    E=&\frac{1}{2}v^{2}-\frac{1}{r}+\frac{\lambda}{2}\left[\frac{3}{4}(1-3\nu)v^{4}\parenthnewln+\frac{1}{r}\left((3+\nu)v^{2}+\nu\frac{(\mathbf{rv})^{2}}{r^{2}}+\frac{1}{r}\right)\right],\\
    \mathbf{J}=&\mathbf{r}\times\mathbf{v}\left[1+\lambda\left(\frac{1}{2}(1-3\nu)v^{2}+(3+\nu)\frac{1}{r}\right)\right].
  \end{split}
  \label{eq:en_impmom}
\end{equation}
This implies that the motion reduces to an effectively one-dimensional bounded problem that in turn leads to a periodic time dependence for all the variables. One can also derive the dimensionless form of the period $P$ between two consecutive pericentre passages \citep{Dam1}. The pericentre advance and $P$ are given as
\begin{equation}
  \begin{split}
    \Delta\phi=6\pi\frac{\lambda}{1-e_{c}^{2}},\mbox{\hspace{3mm}}
    P=\frac{2\pi}{(-2E)^{3/2}}\left[1+\frac{1}{4}(15-\nu)\frac{\lambda}{2}\right].
  \end{split}
  \label{eq:dphi_p}
\end{equation}
The classical eccentricity is obtained from $E_{c}$ and the classical angular momentum $J_{c}$ as
\begin{equation}
    e_{c}=\left[1-J_{c}^{2}\right]^{1/2}.
    \label{eq:a_e}
\end{equation}
The relativistic orbital elements are then \citep{Dam1}
\begin{equation}
  \begin{split}
    a(\lambda)=&-\frac{1}{2E}-(\nu-7)\frac{\lambda}{4}\lambda,\\
    e(\lambda)=&\left[1+2EJ^{2}-\lambda\left((\nu-6)-\left(\frac{5}{4}\nu-\frac{15}{2}\right)J_{c}^{2}\right)\right]^{1/2}.
  \end{split}
  \label{eq:a_e_lambda}
\end{equation}
Since the masses of primaries are equal, we can set the value of $\nu=1/4$. 
The solution of Eqs.~(\ref{eq:pn}) provides the time dependent driving for the relativistic Sitnikov problem.

\subsection{The relativistic Sitnikov problem}

To obtain the equation of the relativistic Sitnikov problem (RSP) we use the Lagrangian of the post-Newtonian 3-body system \citep{Lan}.  The final form of the equation is (for details see the Appendix.):
\begin{equation}
  \begin{split}
    \ddot{z}=-\frac{z}{\rho^3}+\lambda&\left[\frac{5}{4}\frac{z}{r\rho^3}+\frac{16}{2}\frac{z}{\rho^4}
      -\frac{v^{2}z}{\rho^3}\parenthnewln+6\frac{\dot{z}^{2}z}{\rho^3}
      +\frac{3}{2}\frac{(\mathbf{vr})\dot{z}}{\rho^3}+\frac{3}{16}\frac{(\mathbf{vr})^{2}z}{\rho^5}\right],
  \end{split}
\label{eq:rsp}
\end{equation}
here $z$ and $\dot z$ are the dimensionless position and the velocity of the test particle, respectively, and $\rho=\sqrt{z^{2}+r^{2}/4}.$ The additional terms in the bracket on the right hand site of Eq.~(\ref{eq:rsp}) describe the relativistic effects on Sitnikov's motion. 

\section{Numerical results}

As initial conditions to the two-body problem we take the initial conditions at the pericentre in the form $\mathbf{r_{peric}}=(1-e_{c};0),$ $\mathbf{v}=\left(0;((1+e_c)/(1-e_c))^{1/2}\right).$ We shall fix the classical eccentricity to be $e_{c}=0.2$ $(J_{c}=0.9797).$
The total energy is then
\begin{equation}
  \begin{split}
E=-\frac{1}{2}+\frac{\lambda}{32}\frac{71+58e_c+3e_c^2}{(1-e_c)^2}=-0.5+4.04\lambda.
  \end{split}
\label{eq:en_tot}
\end{equation}
The period $P$ up to first order in $\lambda$ is therefore
\begin{equation}
  \begin{split}
P=2\pi(1+14\lambda).
  \end{split}
\label{eq:rel_per}
\end{equation}

\begin{figure}
  \includegraphics[width=80mm]{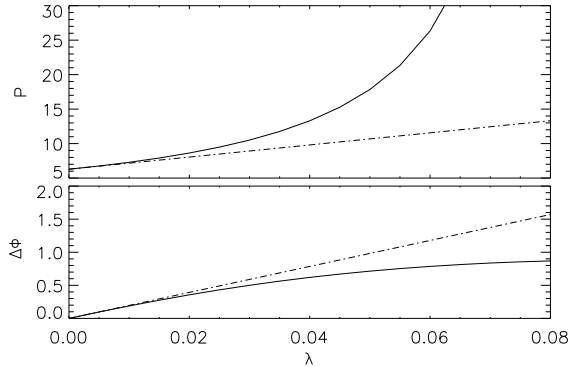}
  \caption{Analytical and numerical solutions of the relativistic two-body problem. Upper panel: Period $P.$ solid line -- numerical solution, dash-dotted -- analytical solution given by Eq.~(\ref{eq:rel_per}). Lower panel: Pericentre advance, solid line -- numerical solution, dash-dotted -- analytical solution given by Eq.~(\ref{eq:dphi_p}).}
  \label{fig:num_an}
\end{figure}

Our numerical investigations show that the post-Newtonian approximation is valid between $0$ and some $\lambda_{c}.$ We define the critical $\lambda_{c}$ as a value where the numerical results of the two-body problem differ from the analytical results (\ref{eq:dphi_p}) by about 10 per cent. Although the equations of motion (\ref{eq:pn}) are valid up to first order in $\lambda,$ they are nonlinear equations and can provide results which are higher order in $\lambda$ or $v^{2}/c^{2}.$ There is thus a threshold beyond which the numerical solution no longer holds. Figure~\ref{fig:num_an} indicates that $\lambda_{c}\approx0.035.$ 

Due to the time periodic driving one can investigate the structure of the three dimensional phase space via stroboscopic or Poincar\'e maps, like in the non-relativistic case.

\subsection{Phase space structure}

\begin{figure}
  \includegraphics[width=80mm]{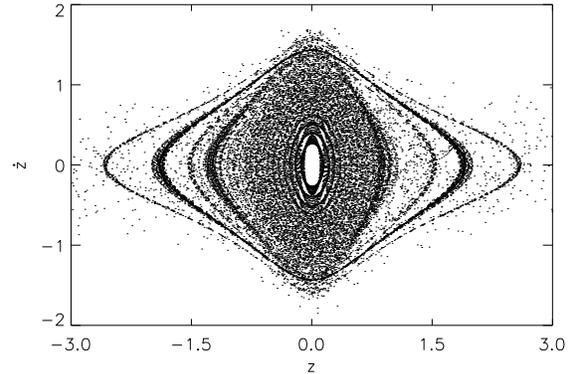}
  \caption{Stroboscopic phase portrait of the RSP ($\lambda=0.005$) taken with the classical period of the two-body problem. Initial conditions are taken from the interval $0.05\leq z\leq 3.0$ ($\Delta z=0.05$) and $\dot z=0.$}
  \label{fig:rel_str}
\end{figure}

Along with the gravitational radius the orbital period changes, as expressed by Eq.~(\ref{eq:rel_per}). Therefore, if data are stored corresponding to the Keplerian orbital period, we obtain a confused phase portrait. Looking at Figure~\ref{fig:rel_str}, one cannot distinguish islands or chaotic bands, we see just "fuzzy" curves and sparse points everywhere. In order to get a ''transparent'' phase portrait, the new period, $P,$ of the primaries' revolution or a redefined Poincar\'e map is needed. In this paper figures are plotted corresponding to the Poincar\'e map taken at $r=r_{peric},$ when the primaries are in the pericentre. 

\begin{figure}
  \includegraphics[width=80mm]{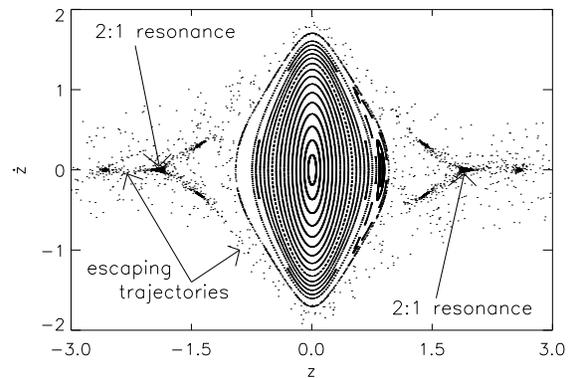}
  \caption{Phase portrait of the RSP at $\lambda=0.005$ taken at the pericentre passage of the primaries. Initial conditions are the same as in Fig.~\ref{fig:rel_str}.}
  \label{fig:rel_pss}
\end{figure}

\begin{figure}
  \includegraphics[width=80mm]{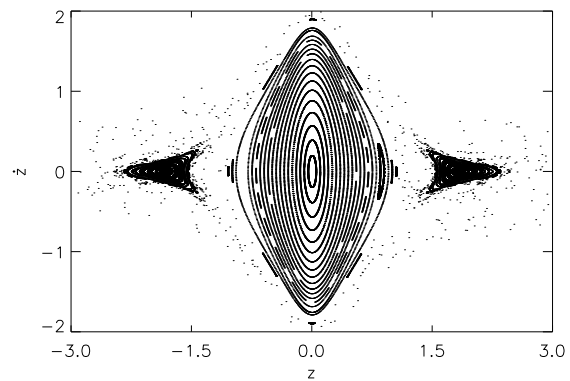}
  \caption{Phase space portrait of the classical Sitnikov problem at $e=e_c=0.2,$ taken at times $2\pi,4\pi,6\pi,\dots$ Initial conditions are the same as in Fig.~\ref{fig:rel_str}.}
  \label{fig:class}
\end{figure}

Figure~\ref{fig:rel_pss} shows this Poincar\'e map that allows us to investigate the phase space as usual. One can indeed see the pattern typical of conservative dynamics.\footnote{For simplicity, instead of the canonically conjugated $(z,p_{z})$ coordinates we use the traditional $(z,\dot{z})$ coordinates, although it makes the Poincar\'e map not exactly area preserving. This map is smoothly conjugated to the area preserving one.} By comparing the relativistic (Fig.~\ref{fig:rel_pss}) and the classical phase portraits (Fig.~\ref{fig:class}), the structure is different but the Hamiltonian characteristics remain.

The RSP has one new parameter, the gravitational radius $\lambda$. The qualitative features of the phase space depends then on $\lambda$ only. One can see from Fig.~\ref{fig:rel_pss} and \ref{fig:class} that even a small $\lambda$ can change the phase portrait dramatically. For $\lambda=0.005$ the main difference to the classical case is the modified surroundings of the 2:1 resonance. The central region is similar to the classical case, invariant tori are situated around the stable origin. However, the main islands corresponding to the 2:1 resonance are changed. The irregularly scattered points around the islands of stability represent the trajectories that can escape the system sooner or later.

\begin{figure*}
  \includegraphics[width=175mm]{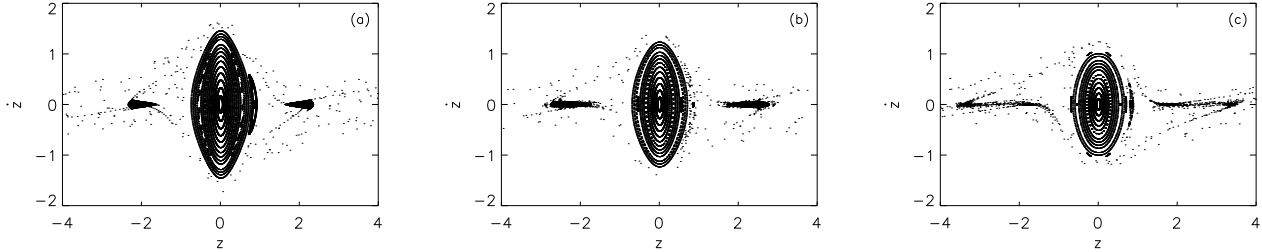}
  \caption{Phase portraits for different gravitational radii (a):$\lambda=0.015$, (b):$\lambda=0.025$, (c):$\lambda=0.035.$ The central stable region becomes smaller from left to right. Moreover, the different shape of the island of the 2:1 resonance shows the sensitivity to $\lambda.$ Panel (c) exhibits  a phase space section after a bifurcation at $\lambda=0.032.$}
  \label{fig:diff_lambda}
\end{figure*}
\begin{figure*}
  \includegraphics[width=175mm]{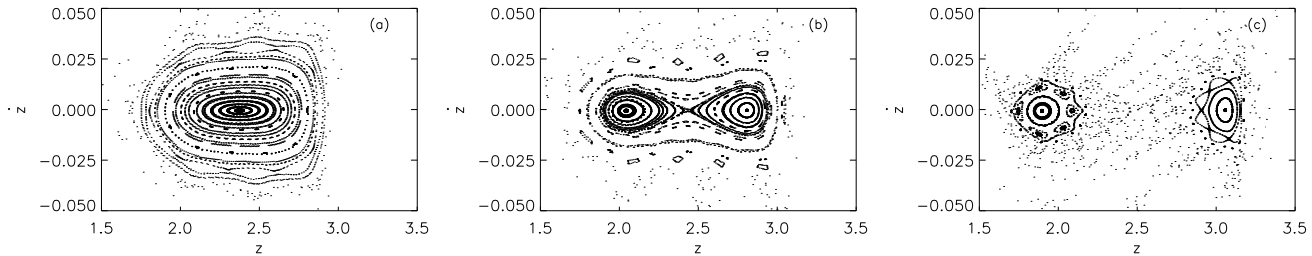}
  \caption{Bifurcation due to changing the gravitational radius about $\lambda\approx 0.03$ Panel (a)-(c): $\lambda=0.028,$ $\lambda=0.03,$ $\lambda=0.032.$ The originally stable periodic orbit (2:1 resonance) becomes unstable and two new stable fixed points appear in the phase portrait.}
  \label{fig:bifurc}
\end{figure*}
Figure~\ref{fig:diff_lambda} shows three phase portraits for different values of the gravitational radius $\lambda.$ 
One can draw several conclusions from these plots. First, it is evident that the central stable region is shrinking when $\lambda$ becomes larger. It means, if the perturbation is larger, the domain of the ordered motion is smaller. Second, the size of the island of the 2:1 resonance becomes larger along the direction parallel to the $z$ axis and smaller perpendicular to it. During this process, the perimeter of the island increases resulting in a longer island chain around the last KAM-torus. Consequently, there are longer Cantori which the scattered trajectories can stick to during the scattering process. We will see this in the next Section.

We have identified a bifurcation at $\lambda\approx0.03.$ The qualitative change due to the increasing gravitational radius is well-seen in Figure~\ref{fig:bifurc}. 
Panel (a) shows the right island of the 2:1 resonance with one stable elliptic fixed point on the phase portrait sitting in the middle of the regular island. However, for larger $\lambda$ (panel (b) and (c)) the stable periodic orbit becomes unstable and two new elliptic fixed points appear to the left and right. 

\begin{figure}
  \includegraphics[width=80mm]{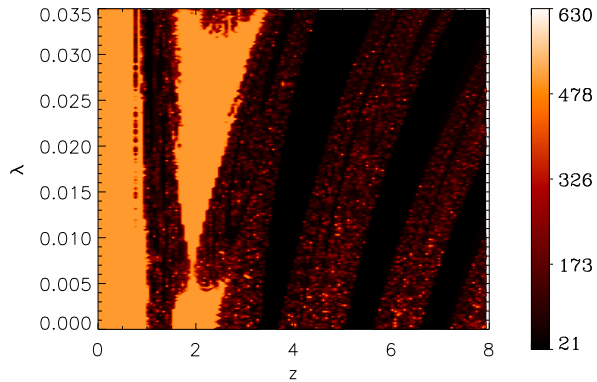}
  \caption{Escape times of the trajectories that originate from $(z,\dot z=0)$ on the $(\lambda,z)$ parameter plane. When the test-particle has positive energy, we store the integration time as the escape time of the orbit. This contour map was calculated over 100 periods of primaries (i.e. roughly 628 time units, see the color bar on the right hand side). The central stable region and the right island of the 2:1 resonance (around $z\approx 2.0$) are plotted in light gray (orange on-line). These trajectories belong to periodic or quasi-periodic orbits and never escape. However, there are other regions far from the stable islands where the escape times are higher than many orbital periods of primaries. These parts of the phase plane contain the stable manifold of the chaotic saddle.}
\label{fig:z_lambda}
\end{figure}

In order to show another picture about how the phase space changes with $\lambda,$ we plotted the escape times in a contour plot (Figure~\ref{fig:z_lambda}). 
If the mechanical energy of the test-particle becomes positive, the trajectory never returns to the primaries' plane, i.e. the test-particle leaves the system. In Figure~\ref{fig:z_lambda} different colors represent different escape times in the $(\lambda,z)$ plane. The contour plot was made as follows. We have chosen 250 initial conditions along the $z$-axis in $[0;8]$ with initial velocity zero, and 500 values of $\lambda$ from the interval $[0;0.035].$ The escape times i.e. the time needed to reach the state where the energy of the particle becomes positive, were computed at the grid points of the $(\lambda,z)$ lattice and plotted with different colors. Figure~\ref{fig:z_lambda} allows us to see the evolution of the extent of the stable islands (where the lifetime is maximal) and filamentary structures. One can see a pitchfork bifurcations when $\lambda$ is growing at $\lambda\approx0.032.$ Beyond the value $\lambda\approx0.012$ the island of 1:1 resonance ($z\approx0.75$) becomes separated from the main central stable region. In other words, trajectories between the resonant island and the central invariant curves may escape to the infinity (Fig.~\ref{fig:z_lambda} and Fig.~\ref{fig:diff_lambda}a). The bright colored filaments with higher escape times correspond to a fractal set representing the stable manifold of a chaotic saddle existing far from the stability islands \citep{Kov2}. Trajectories originating from these initial conditions can spend very long time around the primaries' plane before leaving the system. 

We can say that the parameter $\lambda$ plays a similar role in the
RSP as the eccentricity in the classical case. Varying the
gravitational radius beyond a fixed eccentricity, we obtain
qualitative changes in the phase space as is common in Hamiltonian
dynamics. A complete picture of the phase space of the classical
Sitnikov problem was published in \citet{Dvo2} where escape times show
a structure similar to that of Fig.~\ref{fig:z_lambda}.

\subsection{Chaotic scattering}
In the previous Section we have seen that there are initial conditions which correspond to long life times. Transient chaos appears as a scattering process in conservative dynamics \citep{Eck,Jun,Ble2}. In our example the test particle comes close to the primaries' plane and makes several oscillations before escaping. One can ask where the long-lived trajectories are in the phase space. In order to answer this question, a large number of points is distributed uniformly in the phase space and their evolution in time is followed. We are interested in non-escaping trajectories in a preselected region. Before the trajectories leave the system they draw out a well-defined fractal set in surfaces of sections \citep{Ott}. The invariant object in the phase space responsible for the transient chaotic behavior is the chaotic saddle. They characterize the dynamics in a way chaotic bands characterize permanent chaos. It was also shown that these invariant saddles have two different parts. One of them, the hyperbolic part, is responsible for short lifetimes, the other one, the non-hyperbolic part, is situated close to the border of the KAM-tori and is associated with the sticky orbits \citep{TT,Alt}.

\subsubsection{Short time escape}
\begin{figure}
  \includegraphics[width=80mm]{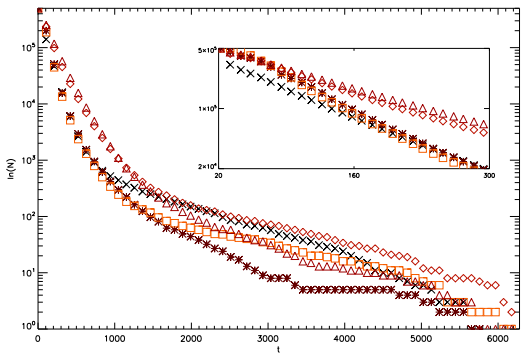}
  \caption{Number $N(t)$ of non-escaped trajectories from a preselected box ($-10\leq z\leq 10$) in the phase space. Different marks represent different values of $\lambda:$ triangles - 0; diamonds - 0.005; crosses - 0.015; asterisks - 0.025; squares - 0.035. The escape rates can be obtained from the slope of the fitted lines on log-lin plot, see the inset on the right. The corresponding escape rates are shown in Table~\ref{tab:1}. Initial conditions: $6\leq z \leq 6.8$, $|\dot z|\leq 0.1.$}
  \label{fig:exp}
\end{figure}

In truly hyperbolic systems the number of non-escaping trajectories decreases exponentially \citep{Kan}. However, in the Sitnikov and also in the relativistic Sitnikov problem we have regular islands in the phase portraits, i.e the phase space is mixed. There are both hyperbolic and non-hyperbolic parts present. In this case the decrease of the survivors follows the exponential rule only for shorter times, as shown in Figure~\ref{fig:exp}.  

If we choose the initial conditions far from regular islands (e.g. $6\leq z\leq 6.8$ and $-0.1\leq \dot z \leq 0.1$), the dynamics can be considered to be hyperbolic. Figure~\ref{fig:exp} shows the number of non-escaped trajectories for different gravitational radii. One can see that the first segments of the curves follow different straight lines in the log-lin plot, $N(t)\sim e^{-\kappa t}$. The escape rate, $\kappa,$ whose inverse tells us the average lifetime of chaos, can be obtained from the slope of these lines. The inset shows that the slopes are not equal. For various $\lambda$ we get different escape rates, see Table~\ref{tab:1}. 

\begin{table}
  \centering
  \caption{Escape rates and average lifetimes. The greater $\lambda$ the shorter the average lifetime of chaos. The function $\kappa(\lambda)$ is close to be linear.}
  \begin{tabular}{@{}lccc@{}}
    \hline
    $\lambda$&escape rate,$\kappa$& escape time, $1/\kappa$&errors\\
    \hline
    $0.0$&$0.007$&$142.8$&$\pm1.4\times10^{-5}$\\
    $0.005$&$0.008$&$125.0$&$\pm1.5\times10^{-5}$\\
    $0.015$&$0.010$&$92.6$&$\pm2\times10^{-5}$\\
    $0.025$&$0.012$&$80.6$&$\pm3.8\times10^{-5}$\\
    $0.035$&$0.013$&$75.8$&$\pm4.7\times10^{-5}$\\
    \hline
    \label{tab:1}
  \end{tabular}
\end{table}

\begin{figure*}
  \includegraphics[width=175mm]{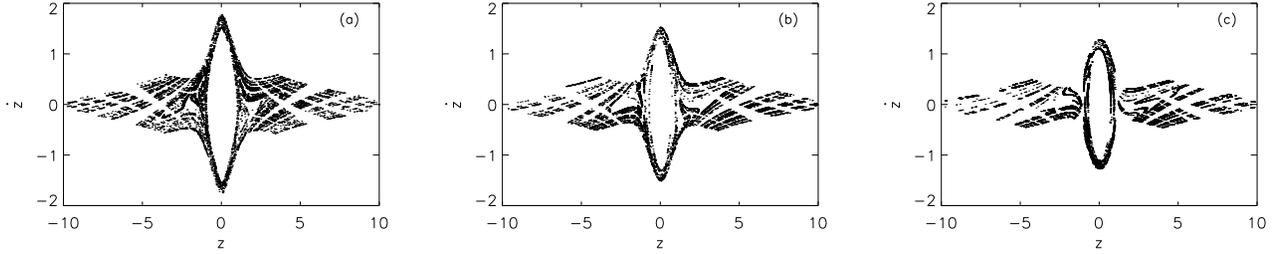}
  \caption{Chaotic saddles for different gravitational radii, (a) $\lambda=0.015$; (b) $\lambda=0.025$; (c) $\lambda=0.035.$ The structure of the saddle far from the ordered regions is similar in each panel, that of the so-called double fractal sets. The shape of the saddle changes more dramatically close to the stability islands. It is evident that the tori do not belong to the saddle. In the numerical simulations $N_{0}=5\cdot 10^{5}.$ The integration time $t_0$ used in the algorithm is $t_0=150$ for (a) and $t_0=180$ for (b),(c).}
\label{fig:saddles}
\end{figure*}

The chaotic saddle responsible for the finite time chaotic motion has a double-fractal structure. One can consider this object as the union of all the hyperbolic unstable periodic orbits and the intersections of their stable and unstable manifolds \citep{TT}. In other words, the scattered test-particle jumps randomly on the saddle before leaving it. One can see the numerically determined saddles for various gravitational radii in Figure~\ref{fig:saddles}. If the number $N_{0}$ of initial conditions is large enough, the implemented method \citep{TT} allows us to visualize the saddle itself. We suppose that the initial point of a trajectory lies close to the stable manifold of the chaotic saddle and we follow the evolution of this point forward in time. After some iteration it must be in the vicinity of the unstable manifold of the saddle. A suitable integration time $t_0$ can be determined from the average lifetime, $1/\kappa,$ of chaos. We have chosen $t_0\approx (2-4)/\kappa.$ Consequently, the mid-point taken at $t\approx t_0/2$ of the trajectory should be close to the saddle. In order to generate the chaotic saddles of Fig~\ref{fig:saddles}, we stored the mid-points of the trajectories that do not escape a preselected box ($|z|\leq 10$ $|\dot z|\leq 2$) in time $t_0.$

The size of the chaotic saddle decreases in Fig.~\ref{fig:saddles} when $\lambda$ is growing and the scenario in Figure~\ref{fig:saddles} is similar to that in Fig.~\ref{fig:diff_lambda}. The double Cantor structure is dominant but empty holes appear at the sites of regular islands. This is the consequence of the quasi-periodic motion on tori which is permanent and certainly not chaotic, but can be arbitrarily close to the chaotic saddle.

\subsubsection{Long time escape -- stickiness}
For longer times the number of non-escaping trajectories does not follow the exponential decay. Instead, one observes a power-law decay which is slower than the exponential one, $N(t)\sim t^{-\sigma}.$ It is an established fact that trajectories which come close to the outer border of the stability islands may stick to them through the debris of the previously destroyed KAM curves. A geometrical consequence of the stickiness effect in the phase space is the denser saddle structure. In other words, one can observe the remnants of previously destroyed KAM-tori, the so called Cantori, around the stability islands, and this part can be identified as the non-hyperbolic part of the chaotic saddle.

\begin{figure}
  \includegraphics[width=80mm]{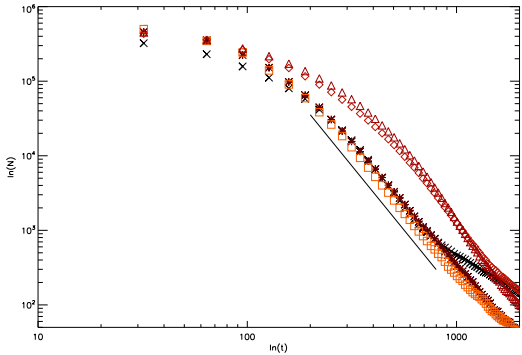}
  \caption{The log-log plot of the number $N(t)$ of non-escaped trajectories. Power-law decay holds in the non-hyperbolic part of the saddle. Trajectories that can come close to the outer KAM-tori may spend very long time around them. The exponent in the power-law decay describes the escape rate of these sticky trajectories. The value of $\sigma$ is not as sensitive as $\kappa$ but our results show that a small fluctuation can be detected related to the size of Cantori surrounding the quasi-periodic regions. The parameters and notation are the same as in Fig.~\ref{fig:exp}.}
\label{fig:power_law}
\end{figure}

Our numerical investigations support the theoretical results, namely, exponent $\sigma$ does not depend on the parameters of the dynamical systems. From fitting straight lines to the points between 500 and 1500 in Figure~\ref{fig:power_law}, we find $\sigma\approx3.6$ irrespective of $\lambda.$


\subsubsection{Basins of escape}
In general, when the test-particle escapes the system, i.e. never comes back, it approaches infinity. Therefore, although in conservative systems there are no attractors, one can consider infinity as an attractor of the system. In other words, beyond the escape energy, infinity behaves as an attractor for those trajectories which leave the system. 

\begin{figure}
  \includegraphics[width=80mm]{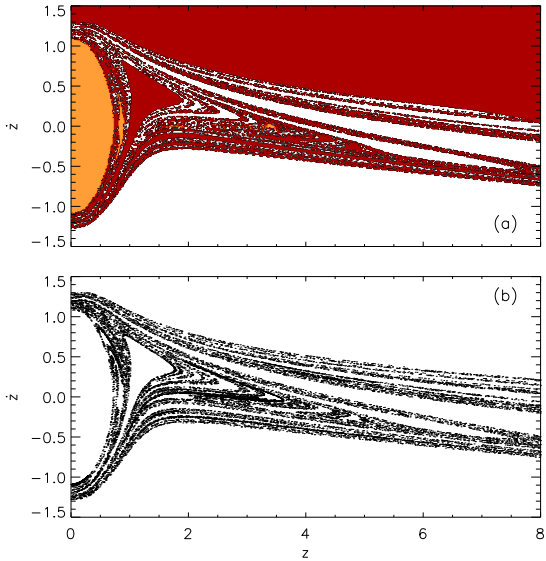}
  \caption{Fractal patterns for $\lambda=0.035.$ (a) Basin boundary. A large number $N_{0}=1.6\cdot 10^{5}$ of points was integrated forward to see which escape route they chose. The dark gray[in red on-line](white) region represents initial conditions leading to an escape to minus(plus) infinity. Light gray [orange on-line] marks the ordered motion inside the regular islands. The very complex structure of the basin boundary indicates chaotic motion. Grid size: $0\leq z\leq 8,$ $\Delta z=0.008;$ $-2\leq \dot z\leq 2,$ $\Delta \dot z=0.004;$  (b) Stable manifold of the chaotic saddle in the relativistic Sitnikov problem. One can see that the filamentary structure of the stable manifold and the fractal basin boundary of the escapes are identical as it is well-known from dissipative systems.}
\label{fig:basins}
\end{figure}

Initializing many initial conditions in a fine rectangular grid one can identify the basins of escape, Figure~\ref{fig:basins}. In the relativistic Sitnikov problem the test-particle may leave the system upward or downward from the primaries' plane depending on the initial conditions of the trajectory; two basins of escape can be identify, the basins of $+\infty$ and $-\infty,$ respectively. Figure~\ref{fig:basins}a shows these basins. The white region contains initial conditions that provides the orbits escaping the system upward, points marked with dark gray color correspond to the orbits leaving the system downward. The third part of the phase portrait (light gray) belongs to regular islands the trajectories never escape from. Considering only the basins of escape one can see the very complex structure of the boundary. The fine-scale structure of the boundary shows that it is not a simple curve, rather a fractal set. Fractality is in general, a result of chaotic motion \citep{McD,Ott}.

Let us consider the analogy with dissipative systems where the stable manifold of hyperbolic unstable fixed points provide the boundary of the basin of attraction. Figure~\ref{fig:basins}b shows the stable manifold of the chaotic saddle, i.e. the initial conditions of those trajectories which remain for very long time in the system. The correspondence is evident between panel (a) and (b). We point out that the fractal basin boundary of escape in open Hamiltonian systems, which is a result of chaotic scattering, is identical with the stable manifold of the chaotic saddle responsible for transient chaos. (See also \citet{TT,Ern})

\section{Conclusions}
In this work we have investigated the relativistic Sitnikov problem numerically. The motivation was to show how the structure of the phase plane changes under the influence of general relativity. The model contains the first post-Newtonian relativistic corrections, derived from the leading order relativistic Lagrangian. Besides the eccentricity of primaries the gravitational radius is the new parameter of the system that is related to the pericentre shift of the two large bodies.   

The calculations show that the problem remains a driven system but the new driving period corresponds to the relativistic orbital period of the binaries. Therefore a new stroboscopic or Poincar\'e map was required to visualize correctly the phase space structure. We found that the Poincar\'e section exhibits the well-known picture of Hamiltonian chaos. Moreover, changing the parameter $\lambda,$ the phase portrait's structure shows qualitative changes. We have pointed out the shrinking of the central regular domain when the gravitational radius becomes larger and simultaneously several bifurcations occurred to the 2:1 resonance. In order to investigate the chaotic scattering through escapes, we have integrated a large number of initial conditions. Two different type of escapes, short and long time escape, were distinguished to identify the dual structure of the chaotic saddle. Short time escapes belong to the hyperbolic part of the saddle, the number of non-escaping trajectories from a preselected region decreases exponentially. During the scattering process such trajectories draw out the double fractal set of the chaotic saddle in the phase space. Escape rates corresponding to different value of gravitational radii show a nearly linear increase with $\lambda.$ The number of trajectories with longer life time follows a power-law decay. Such long time escape characterizes the so-called sticky orbits which may spend very long time in Cantori around the stability islands. The exponent describing the leakage of sticky trajectories seems to not depend on $\lambda.$ From a more general point of view, our results demonstrate that transient chaos and chaotic scattering are {\it robust} phenomena in celestial mechanics as weak relativistic effects are unable to destroy them. Perturbations change the characteristic numbers, at
most. 

\section*{Acknowledgments}
This work was supported by the Hungarian Science Foundation under Grant no. OTKA NK72037. The project is also supported by the European Union and in co-financed by the European Social Fund (grant agreement no. TAMOP 4.2.1./B-09/KMR-2010-0003).

\appendix

\section[]{Equation of motion for the RSP}
We consider the motion of three gravitating masses in the first post-Newtonian
approximation of general relativity. 
The first and second particles have the same mass $M$, opposite velocities,
and they revolve in the $x-y$ plane around the centre of mass which is at the origin.
Their separation and relative velocity is denote by $\mathbf{r}$ and $\mathbf{ v},$ respectively. 
In other terms,
\begin{eqnarray} 
m_1=m_2=M,\hspace{3mm}
\mathbf{ r}_1=-\mathbf{ r}_2=\frac{1}{2}\mathbf{ r},\hspace{3mm}
\mathbf{ v}_1=-\mathbf{ v}_2=\frac{1}{2}\mathbf{ v}.
\end{eqnarray}
The mass $m$
of the third particle is negligible compared to $M$. This third particle moves
along the $z$ axis. Let us denote
$|\mathbf{r}_1-\mathbf{r}_3|=|\mathbf{r}_2-\mathbf{r}_3|$ by $\rho$. Clearly,
\begin{eqnarray} 
\rho=\sqrt{\left(\frac{r}{2}\right)^2+z^2}.
\end{eqnarray}
The Lagrangian of the third
particle can be derived from the three-particle Lagrangian~\citep{Lan} by considering $\mathbf{ r}$ and $\mathbf{ v}$ as given
functions of the time. We get (by omitting full time derivatives)
\begin{eqnarray} 
  L&=&\frac{m}{2}\dot{z}^2+\frac{3kmM}{2c^2\rho}\left(\frac{1}{2}v^2+2\dot{z}^2\right)+\frac{m\dot{z}^4}{8c^2}+\frac{2kmM}{\rho}\nonumber\\
  &+&\frac{kmM}{4c^2\rho^3}(\mathbf{vr})z\dot{z}-\frac{2k^2mM^2}{c^2r\rho}-\frac{k^2mM(m+2M)}{c^2\rho^2},
\end{eqnarray}
or, since $m\ll M$,
\begin{eqnarray} 
  L&=&\frac{m}{2}\dot{z}^2+\frac{3}{4}\frac{kmM}{c^2\rho}v^2+3\frac{kmM}{c^2\rho}\dot{z}^2+\frac{1}{8}\frac{m\dot{z}^4}{c^2}+2\frac{kmM}{\rho}\nonumber\\
  &+&\frac{1}{4}\frac{kmM}{c^2\rho^3}(\mathbf{vr})z\dot{z}-2\frac{k^2mM^2}{c^2r\rho}-2\frac{k^2mM^2}{c^2\rho^2}.
\end{eqnarray}
The partial derivatives of $L$ are:
\begin{eqnarray} 
p_z=\frac{\partial L}{\partial \dot{z}}=m\dot{z}+6\frac{kmM}{c^2\rho}\dot{z}+\frac{1}{2}\frac{m\dot{z}^3}{c^2}+\frac{1}{4}\frac{kmM}{c^2\rho^3}(\mathbf{vr})z,
\end{eqnarray}

\begin{eqnarray} 
  \frac{\partial L}{\partial z}&=&-\frac{3}{4}\frac{kmM}{c^2\rho^3}v^2z-3\frac{kmM}{c^2\rho^3}\dot{z}^2z-2\frac{kmM}{\rho^3}z\nonumber\\
  &-&\frac{3}{4}\frac{kmM}{c^2\rho^5}(\mathbf{vr})z^2\dot{z}+\frac{1}{4}\frac{kmM}{c^2\rho^3}(\mathbf{vr})\dot{z}+2\frac{k^2mM^2}{c^2r\rho^3}z\nonumber\\
  &+&4\frac{k^2mM^2}{c^2\rho^4}z.
\end{eqnarray}

When deriving the equation of motion, in the correction terms we replace the second derivatives with their zeros order expressions, namely
\begin{eqnarray} 
\dot{\mathbf{v}}=-2\frac{kM}{r^3}\mathbf{r},\hspace{3mm}
\ddot{z}=-2\frac{kM}{\rho^3}z.
\end{eqnarray}
Further, we use the relation
\begin{eqnarray} 
  \dot{\rho}=\frac{1}{4}\frac{\mathbf{vr}}{\rho}+\frac{\dot{z}z}{\rho}
\end{eqnarray}
to obtain
\begin{eqnarray} 
  \ddot{z}&=&-2\frac{kM}{\rho^3}z+16\frac{k^2M^2}{c^2\rho^4}z+6\frac{kM}{c^2\rho^3}\dot{z}^2z+\frac{3}{2}\frac{kM}{c^2\rho^3}(\mathbf{vr})\dot{z}\nonumber\\
  &+&\frac{3}{16}\frac{kM}{c^2\rho^5}(\mathbf{vr})^2z-\frac{kM}{c^2\rho^3}v^2z+\frac{5}{2}\frac{k^2M^2}{c^2r\rho^3}z.
\end{eqnarray} 
After rearranging terms,
\begin{eqnarray} 
  \ddot{z}&=&-2\frac{kM}{\rho^3}z+\frac{5}{2}\frac{k^2M^2}{c^2r\rho^3}z+16\frac{k^2M^2}{c^2\rho^4}z-\frac{kM}{c^2\rho^3}v^2z\nonumber\\
  &+&6\frac{kM}{c^2\rho^3}\dot{z}^2z+\frac{3}{2}\frac{kM}{c^2\rho^3}(\mathbf{vr})\dot{z}+\frac{3}{16}\frac{kM}{c^2\rho^5}(\mathbf{vr})^2z.
\label{eq:rel_sit}
\end{eqnarray} 

In order to obtain the dimensionless equations, first, we take the semi-major axis ($a$) of the Keplerian orbit as unit length. Consequently, the unit of the velocity is \(a/T,\) where \(T\) denotes the time unit. As in Section 2.1 we choose \(T\) as \(a^{3/2}/(k2M)^{1/2}\) corresponding to Kepler's 3rd law (up to a constant \(2\pi\)). Thus, the dimensionless equation from (\ref{eq:rel_sit}) is obtained as Equation~(\ref{eq:rsp}) in the main text.

\bsp

\label{lastpage}

\end{document}